%% file: main.tex
\documentclass{oax}
\usepackage{url}
\usepackage{graphicx}
\graphicspath{{figs/}}
\usepackage{subfigure}
\usepackage{xspace}
\usepackage{hyperref}

\usepackage{tabularx}
\usepackage{listings}
\usepackage{color}

\newcommand{\libcppa}{\texttt{libcppa}\xspace}

\newcommand{\cpp}{C++\xspace}

\newcommand{\erlang}{Erlang\xspace}
\newcommand{\scala}{Scala\xspace}
\newcommand{\akka}{Akka\xspace}

\hypersetup{%
  colorlinks=false,%
  pdfborder = 0 0 0,%
  pdfproducer={},%
  bookmarksnumbered,%
  bookmarksopen=false,%
  pdfstartview=Fit,%
  linkcolor=black,%
  citecolor=black,%
  filecolor=black,%
  menucolor=black,%
  urlcolor=black,%
  plainpages=false,%
  hypertexnames=false%
}

\newcommand{\footnoterecall}[1]{%
  \hyperref[#1]{\footnotemark[\value{#1}]}
}

\definecolor{grey}{rgb}{0.5,0.5,0.5}

\definecolor{blue}{rgb}{0,0,1}
\definecolor{violet}{rgb}{0.5,0,0.5}

\definecolor{midblue}{rgb}{0,0,0.75}

\definecolor{darkred}{rgb}{0.5,0,0}
\definecolor{darkblue}{rgb}{0,0,0.5}
\definecolor{darkgreen}{rgb}{0,0.5,0}

\begin{document}

\clubpenalty=10000
\widowpenalty = 10000

\title{libcppa -- Designing an Actor Semantic for C++11}

\numberofauthors{2}

\author{
\alignauthor
Dominik Charousset\\\affaddr{HAW Hamburg}
\email{dcharousset@acm.org}
\alignauthor
Thomas C. Schmidt\\\affaddr{HAW Hamburg}
\email{t.schmidt@ieee.org}
}

\maketitle

\begin{abstract}
Parallel hardware makes concurrency mandatory for efficient program execution.
However, writing concurrent software is both challenging and error-prone.
C++11 provides standard facilities for multiprogramming, such as atomic operations with acquire/release semantics and RAII mutex locking, but these primitives remain too low-level.
Using them both correctly and efficiently still requires expert knowledge and hand-crafting. 
The actor model replaces implicit communication by sharing with an explicit message passing mechanism.
It applies to concurrency as well as distribution, and a lightweight actor model implementation that schedules all actors in a properly pre-dimensioned thread pool can outperform equivalent thread-based applications.
However, the actor model did not enter the domain of native programming languages yet besides vendor-specific island solutions.
With the open source library \libcppa, we want to combine the ability to build reliable and distributed systems provided by the actor model with the performance and resource-efficiency of C++11.
\end{abstract}

\category{D.3.3}{Programming languages}{Language Constructs and Features}[Concurrent programming structures]

\vspace*{-2mm}

\keywords{C++, actor model, pattern matching}

\input{part1}
\input{part2}
\input{part3}
\input{part4}
\input{part5}

\input{part6}
\input{part7}
\input{part8}
\input{part9}

\bibliographystyle{acm}
\bibliography{own,programming,rfcs}

\end{document}

%% file: part1.tex
\section{Introduction}

The actor model is a formalism describing concurrent entities, ``actors'', that communicate by asynchronous message passing \cite{hbs-umafa-73}.
An actor can send messages to addresses of other actors and can create new actors.
Actors do not share state and are executed concurrently.

Because Actors are self-contained and do not rely on shared resources, race conditions are avoided by design.
The message passing communication style also allows network transparency and thus applies to both concurrency, if actors run on the same host on different cores/processors, and distribution, if actors run on different hosts connected via the network.

The actor model inspired several implementations in computer science, either as basis for languages like Erlang or as libraries/frameworks such as the Scala Actor Library to ease development of concurrent software.
However, low-level primitives are still widely used in \cpp.
The standardization committee added threading facilities and synchronization primitives to the C++11 standard \cite{iso-cpp-11}, but did neither address the general issues of multiprogramming, nor distribution.

This work targets at concurrency and distribution in \cpp.
We design and implement a library for \cpp, called \libcppa, with an internal Domain-Specific Language (DSL) approach to provide high-level abstraction.

In the remainder, we discuss the actor model and related work in Section \ref{Sec:Background}.
Section \ref{Sec:ActorSemantic} details design decisions of our internal domain-specific language approach.
The pattern matching implementation of \libcppa, which is an important ingredient in our library design, is presented in Section \ref{Sec:PatternMatching}.
Section \ref{Sec:MessagePassing} discusses message passing in \libcppa.
In Section \ref{Sec:Mailbox}, we present our algorithm for message queueing, which is a critical component of actor systems.
Section \ref{Sec:PerformanceEvaluation} evaluates the performance characteristics of \libcppa.
Section \ref{Sec:Limitations} discusses limitations we had to face in C++11 when implementing \libcppa.
Finally, we report on future work in Section \ref{Sec:FutureWork}.

%% file: part2.tex
\section{Background and Related Work}
\label{Sec:Background}

Concurrent software components often need to share or exchange data.
In shared memory environments, this communication happens implicitly via shared states. However, unsynchronized parallel access to shared memory segments easily leads to  erroneous behavior caused by race conditions.
Synchronization protocols are usually built on low-level primitives such as locks and condition variables that prevent parallel or interleaved execution of so called \textit{critical sections}.
This is inherently error-prone and correctly implementing critical sections requires expert knowledge, about compiler optimizations and out-of-order execution for instance \cite{ma-cpdcl-04}.
Still, designing  a suitable locking strategy  is not the only challenge developers face on parallel hardware.
Applications with good performance on a uniprocessor machine may experience a performance degradation on  multiprocessor platforms, e.g., due to \textit{false sharing}.
False sharing occurs whenever two or more processors are repeatedly writing into a memory region that is mapped to more than one processor cache.
This mutually invalidates the caches, even if the processors are accessing distinct data, and seriously slows down execution \cite{tlh-fsslm-94}.
An object that needs to support parallel access by a group of participants should not rely on locking, since a critical section is a performance bottleneck per se.
Lock- and wait-free algorithms \cite{h-wfs-91} scale in a multiprocessor environment, but are significantly more complex to implement and verify \cite{dglm-fvplf-04}.

\subsection{The Actor Model}

Higher-level abstractions following message passing or transactional memory paradigms were developed to establish programming models that are free of race conditions, and thus do not require synchronization of readers and writers.
Transactional memory can be implemented in hardware \cite{hm-tmasl-93} or software \cite{st-stm-95}, and can be seamlessly integrated into new programming languages, as the example of Clojure \cite{pc-hb-12} illustrates.
Transactional memory scales for multiprocessor machines and has been implemented for tightly-coupled distributed systems such as clusters \cite{bac-stmls-08}, but it applies neither to communication between heterogeneous hardware components, nor to widely distributed applications.
Message passing, on the other hand, has proven to scale in multiprocessor environments as well as in both tightly and loosely coupled distributed computing.
In the high-performance domain, message passing systems based on the MPI are used for decades \cite{glds-hppim-96}.

The actor model  \cite{hbs-umafa-73} is build on the message passing paradigm, but raises the level of abstraction even further.
It not only describes how software components communicate, but also characterizes the communicating components, the actors themselves.
Based on this model, concurrent and distributed systems can be composed of independent modules \cite{a-amccd-86} that are open for communication with external components \cite{amst-ttac-92}.
In addition, the actor model addresses reliability and fault tolerance in a network-transparent way \cite{a-mrdsp-03}.

\subsection{Message Processing}

The original actor modeling of Agha \cite{a-amccd-86} introduced mailbox-based message processing.
A mailbox is a FIFO ordered message buffer that is only readable by the actor owning it, while all actors are allowed to enqueue a new message.
Since actors do not share state, the mailbox is the only way of communication among actors. Implementation concepts of  mailbox management divide into two categories.

In the first category, an actor iterates over messages in its mailbox. On each receive call, it begins with the first but is free to skip messages.
This postponing of messages can be automated, if the actor's behavior is defined as a partial function \cite{ho-sautb-09}. As actors can change their behavior in response to a message,
the newly defined behavior may then be applied to previously skipped messages.
A message remains in the mailbox until it is eventually processed and removed as part of its consumption.
Actor systems that are based on this type of message processing include Erlang \cite{a-mrdsp-03} and the Scala Actors library~\cite{ho-sautb-09}.

The second category of actor systems restricts the operations of message processing.
In these approaches, the runtime system invokes a message handler exactly once per message in order of arrival.
Some systems allow changing the message handler, but an untreated message cannot be recaptured later.
Examples of this kind are SALSA \cite{va-pdros-01}, Akka \cite{ti-a-12},
Kilim \cite{sm-kitaj-08}, and Retlang \cite{r-r-10}.

We have followed the first approach, since this allows actors to prioritize messages and to wait for a response prior to returning to the default behavior.
Furthermore, pattern matching has proven useful and very effective to ease definition of partial functions used as message handlers \cite{ho-jmlc-06}.
We provide pattern matching for message handling as a domain-specific language in our system.

\subsection{Fault Propagation}
\label{Sec::Theory::MonitoringOfActors}

Actors achieve fault tolerance by monitoring each other \cite{hbs-umafa-73},
which  is necessary in particular for constructing fault-tolerant distributed systems.
Whenever an actor fails, an exit message is sent to all actors that monitor it.
This monitoring can be bidirectional, and actors with such a strong relation are called \textit{linked}.
Linked actors form a subsystem in which errors are propagated via exit messages.

Based on this mechanism, developers can build subsystems in which all actors are either alive or have collectively failed.
Each subsystem can include one or more actors that survey working actors and re-create failing workers.
In proceeding this way,  hierarchical, fault-tolerant systems such as Erlang's \textit{supervision trees} \cite{a-mrdsp-03} can be built.
Newer implementations of the actor model (e.g., Kilim, Akka, and the Scala
Actors Library) have adopted Erlang's model of error propagation, as it~has proven very effective, elegant and reliable \cite{ntk-edflt-03}.
It was therefore natural for us to adopted Erlang's well-established fault propagation model as well.

%% file: part3.tex
\section{Actor Semantic as Internal DSL}
\label{Sec:ActorSemantic}

Our aim is to add an actor semantic to \cpp that enables developers to build efficient concurrent and distributed software.
Though \libcppa is influenced by functional programming languages such as Erlang and Scala, it should not mimic a functional programming style and provide an API that looks familiar to \cpp developers.

We decided to use an internal Domain-Specific Language (DSL) approach to raise the level of abstraction in \cpp, while requiring as little  glue code as possible.
An essential ingredient of our design is the keyword-like identifier \lstinline^self^.
From a user's pointer of view, it identifies an actor similar to the implicit \lstinline^this^ pointer identifying an object within a member function.

The interface for actors is essentially split in two.
The first interface, called \lstinline^actor^, provides all operations needed for linking, monitoring, and delivering messages.
In \libcppa, actors are always guarded by \lstinline^actor_ptr^, which is a smart pointer type.
The second interface is \texttt{local\_actor}.
It represents the \textit{private} access to an actor, since only the actor itself is allowed to dequeue a message from its mailbox.
The keyword \lstinline^self^ behaves like a pointer of type \lstinline^local_actor*^.

\subsection{Cooperative Scheduling of Actors}

Scalability in the context of multi-core processors requires to split the application logic into many independent tasks that could be executed in parallel.
An actor is a representation of such an independent task.
This makes lightweight creation and destruction of actors mandatory.
It is not feasible to map each actor to its own thread, since creation and destruction of threads is heavyweight.
Thread management relies on system calls and acquires system resources such as thread state in the OS scheduler, stack and signal stack.
Thus, short-living threads do not scale well, since the effort for creation and destruction outweighs the benefit of parallelization.

An ideal way to ensure \textit{fairness} in an actor system requires preemptive scheduling.
A fair system would guarantee that no actor could starve other actors by occupying system resources.
However, unrestricted access to hardware or kernel space is needed to implement preemptive scheduling, since it relies on hardware interrupts to switch between two running tasks.
No operating system can allow unrestricted hardware or kernel space access to a user space application for obvious reasons.

In general, user space schedulers can only implement cooperative scheduling, i.e., execute actors in a thread pool.
But there is some design space to make context switching implicit.
The two operations each actor uses frequently are sending and receiving of messages.
Thus, the library could switch the actor's context back to the scheduler, whenever it sends or receives a message.
An ordinary workflow of an actor is receiving messages in a loop and sending messages either as part of its message processing or after computing  results.
Thus, interrupting an actor during send will probably interrupt an actor during its message processing, while interrupting it during receive seems natural in this workflow.
Furthermore, we need to support interruption during receive since an actor is not allowed to block while its mailbox is empty.
Instead, an actor returns to scheduler and is re-scheduled again after a new message arrives to its mailbox.

Nevertheless, developers can choose to opt-out of the cooperative scheduling and execute an actor in its own thread.
This is particularly useful, whenever an actors needs to call blocking functions and therefore would possibly starve other actors in a cooperative scheduling.

\subsection{Context-Switching Actors}

Context-switching actor use so called \textit{fibers} or \textit{lightweight threads}.
Like a kernel thread, each fiber has its own stack.
However, fibers are scheduled in user space only, and are scheduled on top of kernel threads.
Hence, the ratio between fibers and threads is N:M, as any number of fibers can be cooperatively scheduled on top of kernel threads.

We provide this actor implementation to ease soft migration strategies, where a previously threaded application is ported to actors.
The API for context-switching actors is equal to the API for threaded actors and provides a `blocking' -- at least from the user's point of view -- receive function.
However, this implementation does not scale up to large actor system.
As an example for a current mainstream system: Mac OS X defines the recommended stack size to 131,072 bytes, and the minimally allowed stack size to 32,768 bytes.
Assuming a system with 500,000 actors, one would require a memory usage of \textit{at least} 15 GB of RAM for stack space only.
This would rise up to 61 GB with the recommended
stack size instead in use.
This clearly does not scale well for large systems.
To reduce memory usage and scale up to large systems, we provide a third, event-based approach.

\subsection{Event-Based Actors}
\label{Sec::Design::EventBasedActors}

A common problem of event-based APIs is \textit{inversion of control}.
Event-based APIs usually define a callback that is invoked for each event.
However, callback-based message processing cannot prioritize messages since the callback is invoked by the runtime system usually in arrival order, and thus has a semantic different from our mailbox-based actors using the receive statement.
Therefore, we use a behavior-based API with mailbox-based message processing.

An event-based actor sets its required behavior as a partial function using the \lstinline^become^ function.
This partial function is used until the actor replaces it by invoking \lstinline^become^ again.
Thus, all actor implementations in \libcppa are able to prioritize communication.

\subsection{Emulating the Keyword ``self''}

An actor semantic needs to be consistent.
For that reason, we introduce the identifier \lstinline^self^, which is not allowed to be invalid or to return \lstinline^nullptr^.
Otherwise, it could not be guaranteed that receive or send statements never fail.
This implies that a non-actor caller, i.e., a thread, is converted to an actor if needed.
For instance, when using send and receive functions in \lstinline^main^, the corresponding thread should be converted implicitly to an actor.

Unlike \lstinline^this^, \lstinline^self^ is not limited to a particular scope.
Furthermore, it is not just a pointer, but it needs to perform implicit conversions on demand.
Consequently, \lstinline^self^ requires
a type allowing implicit conversion to
\lstinline^local_actor*^, where the conversion function returns
a thread-local pointer. Our approach shown below uses a
global \lstinline^constexpr^ variable
with a type that behaves like a pointer.

\begin{lstlisting}
class self_type {
  static local_actor* get_impl();
  static void set_impl(local_actor* ptr);
 public:
  constexpr self_type() { }
  inline operator local_actor*() const {
    return get_impl();
  }
  inline local_actor* operator->() const {
    return get_impl();
  }
  inline void set(local_actor* ptr) const {
    set_impl(ptr);
  }
};
namespace { constexpr self_type self; }
\end{lstlisting}

The \lstinline^constexpr^ variable \lstinline^self^ provides access to the implicit conversion operator as well as the dereference operator ``\lstinline^->^''.
From a user's point of view, \lstinline^self^ is not distinguishable from a pointer of type \lstinline^local_actor^.
The static member functions are implemented as follows.

\begin{lstlisting}
thread_local local_actor* t_self = nullptr;
local_actor* self_type::get_impl() {
  if (!t_self) t_self = convert_thread();
  return t_self;
}
void self_type::set_impl(local_actor* ptr) {
  t_self = ptr;
}
\end{lstlisting}

Our approach adds little, if any, overhead to an application. In fact,
\lstinline^self^ is nothing but syntactic sugar and the compiler
could easily optimize away the overhead of using member functions.
A \lstinline^constexpr^ variable does
not cause a dynamic initialization at runtime, why
the global variable \lstinline^self^ does not cause any
overhead since it provides an empty
\lstinline^constexpr^ constructor.
Furthermore, all member functions are declared
\lstinline^inline^, allowing the compiler to replace each
occurrence of a member function with a call to
\lstinline^self_type::get_impl^.
\lstinline^self.set()^ is intended for in-library use only.
The latter is needed to implement cooperative scheduling.

\subsection{Copy-On-Write Tuples}

A message can be send to multiple receivers or forwarded from one actor to another.
In \libcppa, messages always follow call-by-value semantic.
This unburdens the developers of managing tuple lifetimes and keeps the programming model clean and easy to understand.
However, sending a message to multiple actors would require multiple copies of the message, wasting both computing time and memory.
We use a copy-on-write implementation to avoid both unnecessary copies and race conditions.

All messages use the type \lstinline^any_tuple^, which represents a tuple of arbitrary length with dynamically typed elements.
To restore the static type informations, \lstinline^tuple_cast^ can be used, e.g., \lstinline^auto x = tuple_cast<int,int>(tup)^ tries to cast \lstinline^tup^ of type \lstinline^any_tuple^ to a tuple of two integers and returns \lstinline^option<cow_tuple<int,int>>^.
Since the cast returns an option rather than a straight value, the user has to verify the result, just as users of \lstinline^dynamic_cast^ are required to do.
The template class \lstinline^cow_tuple<int,int>^ then provides two access functions, \lstinline^get^ and \lstinline^get_ref^.
Unlike the access for \lstinline^std::tuple^, we do not provide const overloaded functions.
The reason for avoiding this is that non-const access has implicit costs attached to it, because the non-const access causes a deep copy operation if there is more than one reference to the tuple.
Hence, we designed an explicit non-const access function, \lstinline^get_ref^, to prevent users from accidently create unnecessary copies.
Still, users will very seldom, if ever, interact with tuples directly, because the tuple handling is usually hidden by pattern matching, as shown in Section \ref{Sec:PatternMatching}.

\subsection{Spawning Actors}

Actors are created using the function \lstinline^spawn^.
The recommended way to implement both context-switching and thread-mapped actors is to use functors, such as free functions or lambda expressions.
The arguments to the functor are passed to \lstinline^spawn^ as additional arguments.
The optional \lstinline^scheduling_hint^ template parameter of \lstinline^spawn^ decides whether an actor should run in its own thread or use context-switching.
The flag \lstinline^detached^ causes \lstinline^spawn^ to create a thread-mapped actor, whereas \lstinline^scheduled^, the default flag, causes it to create a context-switching actor.
The function \lstinline^spawn^ is used quite similar to \lstinline^std::thread^, as shown in the examples below.

\begin{lstlisting}
#include "cppa/cppa.hpp"
using namespace cppa;

void fun1();
void fun2(int arg1, const std::string& arg2);

class dummy : public event_based_actor {
  // ...
  dummy() { /*...*/ }
  dummy(int i) { /*...*/ }
};

int main() {
  // spawn context-switching actors
  // equal to spawn<scheduled>(fun1)
  auto a1 = spawn(fun1);
  auto a2 = spawn(fun2, 42, "hello actor");
  // spawn a lambda expression
  auto a3 = spawn([]() { /*...*/ });
  auto a4 = spawn([](int) { /*...*/ }, 42);
  // spawn thread-mapped actors
  auto a5 = spawn<detached>(fun1);
  auto a6 = spawn<detached>(/*...*/);
  // spawn actors using class defintions
  auto a7 = spawn<dummy>();
  auto a7 = spawn<dummy>(42);
}
\end{lstlisting}

In general, context-switching and thread-mapped actors are intended to ease migration of existing applications or to implement managing actors on-the-fly using lambda expressions.
Class-based actors usually subtype \lstinline^event_based_actor^.

It is worth noting that \lstinline^spawn(fun, arg0, ...)^ is \textit{not} the same as \lstinline^spawn(std::bind(fun, arg0, ...))^.
The function \lstinline^spawn^ evaluates \lstinline^self^ arguments immediately and forwards them as \lstinline^actor_ptr^ instances.
When using \lstinline^bind^, \lstinline^self^ is evaluated upon function invocation, thus pointing to the spawned actor itself rather than to the actor that created it.

%% file: part4.tex
\section{Pattern Matching in C++}
\label{Sec:PatternMatching}

C++ does not provide pattern matching facilities.
A general pattern matching solution for arbitrary data structures would require a language extension.
Hence, we had decided to restrict our implementation to dynamically typed tuples, to be able to use an internal domain-specific language (DSL) approach.

\subsection{Match Expressions}

A match expression in \libcppa begins with a call to the function \lstinline^on^, which returns an intermediate object providing the member functions \lstinline^when^ and the right shift operator.
The right-hand side of the operator denotes a callback, usually a lambda expression that should be invoked on a match, as shown in the examples below.

\begin{lstlisting}
on<int>() >> [](int i)
on<int,float>() >> [](int i, float f)
on<int,int,int>() >> [](int a, int b, int c)
\end{lstlisting}

The result of such an expression is a partial function that is defined for the types given to \lstinline^on^.
A comma separated list of partial functions results in a single partial function that sequentially evaluates its subfunctions.
A partial function invokes at most one callback, since the evaluation stops at the first match.

\begin{lstlisting}
auto fun = (
  on<int>() >> [](int i) { /*case1*/ },
  on<int>() >> [](int i) {
    // unreachable; case1 always matches first
  }
);
\end{lstlisting}

Due to the C++11 grammar, a list of partial function definitions must be enclosed in brackets if assigned to a variable.
Otherwise, the compiler assumes commas to separate variable definitions.

The function ``\lstinline^on^'' can be used in exactly two ways.
Either with template parameters only or with function parameters only.
The latter version deduces all types from its arguments and matches for both type and value.
The template ``\lstinline^val^'' can assist to match for types.

\begin{lstlisting}
on(42) >> [](int i) { assert(i == 42); }
on("hello world") >> []() {}
on(1, val<int>) >> [](int i) {}
\end{lstlisting}

Callbacks can have less arguments than given to the pattern, but it is only allowed to skip arguments from left to right.

\begin{lstlisting}
on<int,int,float>() >> [](float)
on<int,int,float>() >> [](int, float)
on<int,int,float>() >> [](int, int, float)

// invalid: on<int,int,float>() >> [](int i)
\end{lstlisting}

\subsection{Atoms}
\label{Sec::PatternMatching::Atoms}

Assume an actor provides a mathematical service on integers.
It takes two arguments, performs a predefined operation and returns the result.
It cannot determine an operation, such as \textit{multiply} or \textit{add}, by receiving two operands alone.
Thus, the operation must be encoded into the message.
The Erlang programming language introduced an approach to use non-numerical
constants, so-called \textit{atoms}, which have an unambiguous, special-purpose type and do not have the runtime overhead of string constants.
Atoms are mapped to integer values at compile time in \libcppa.
This mapping is collision-free and invertible, but limits atom literals to ten characters and prohibits special characters.
Legal characters are ``\lstinline[language=C++]^_0-9A-Za-z^'' and whitespaces.
Although user-defined literals are the natural candidate to choose, we have implemented atoms using a \lstinline^constexpr^ function, called \lstinline^atom^, because user-defined literals are not yet available at mainstream compilers.

\begin{lstlisting}
on<atom("add"),int,int>() >> ...
on<atom("multiply"),int,int>() >> ...
\end{lstlisting}

Our implementation can only enforce the length requirement, but cannot enforce valid characters at compile time.
Each invalid character is mapped to the whitespace character, why the assertion \lstinline^atom("!?") != atom("?!")^ is not true.
However, this issue will fade away after user-defined literals become available in mainstream compilers, because it is then possible to raise a compiler error for invalid characters.

\subsection{Reducing Redundancy}

In our previous examples, we always have repeated the types from the left side of a match expression.
To avoid such redundancy, \lstinline^arg_match^ can be used as last argument to the function \lstinline^on^.
This causes the compiler to deduce all further types from the signature of the given callback.

\begin{lstlisting}
on<atom("add"),int,int>() >> [](int a, int b)
on(atom("add"),arg_match) >> [](int a, int b)
\end{lstlisting}

The second example calls \lstinline^on^ without template parameters, but is equal to the first one.
When used, \lstinline^arg_match^ must be passed as last parameter.
If all types should be deduced from the callback signature, \lstinline^on_arg_match^ can be used, which is equal to \lstinline^on(arg_match)^.
Both \lstinline^arg_match^ and \lstinline^on_arg_match^ are \lstinline^constexpr^ variables.

\subsection{Wildcards}
\label{Sec::PatternMatching::Wildcards}

The type \lstinline^anything^ can be used as wildcard to match any number of any types.
A pattern created by \lstinline^on<anything>()^ -- or its alias \lstinline^others()^ -- is useful to define a default ``catch all'' case.
For patterns defined without template parameters, the \lstinline^constexpr^ value \lstinline^any_vals^ can be used as function argument.
The constant \lstinline^any_vals^ is of type \lstinline^anything^ and is nothing but syntactic sugar for defining patterns.

\begin{lstlisting}
on<int,anything>() >> [](int i)
  { /* tuple with int as first element */ },
on(any_vals,arg_match) >> [](int i)
  { /* tuple with int as last element */ },
others() >> [] { /* default handler */ }
\end{lstlisting}

\subsection{Guards}

Guards allow to constrain a given match statement by using placeholders, as the following example illustrates.

\begin{lstlisting}
// contains _x1 - _x9
using namespace cppa::placeholders;

on<int>().when(_x1%2==0) >> []{ /* even */ },
on<int>() >> []{ /* odd */ }
\end{lstlisting}

Guard expressions are a lazy evaluation technique.
The placeholder \lstinline^_x1^ is substituted with the first value of a given tuple.
To reference variables of the enclosing scope, or member variables of the actor itself, we provide two functions designed to be used in guard expressions: \lstinline^gref^ (``guard reference'') and \lstinline^gcall^ (``guard function call'').
The function \lstinline^gref^ creates a reference wrapper that forces lazy evaluation.

\begin{lstlisting}
int val = 42;
// (1) matches if _x1 == 42
on<int>().when(_x1 == val)
// (2) matches if _x1 == val
on<int>().when(_x1 == gref(val))
// (3) matches as long as val == 42
others().when(gref(val) == 42)
// (4) ok, because _x1 forces lazy evaluation
on<int>().when(_x1 == std::ref(val))
// (5) compiler error due to eager evaluation
others().when(std::ref(val) == 42)   
\end{lstlisting}

Statement \texttt{(5)} in the example above is evaluated immediately and returns a boolean instead of a guard expression.
The second function -- \lstinline^gcall^ -- encapsulates a function call.
Its use is similar to \lstinline^std::bind^, but there is also a short version for unary functions: \lstinline^_x1(fun)^ is equal to \lstinline^gcall(fun, _x1)^.

\begin{lstlisting}
typedef std::vector<int> ivec;
auto vsorted = [](const ivec& v) {
  return std::is_sorted(v.begin(), v.end());
};
on<ivec>().when(gcall(vsorted, _x1))
// equal to
on<ivec>().when(_x1(vsorted)))
\end{lstlisting}

\subsection{Projections and Extractors}

Projections perform type conversions or extract data from tuples.
Functors passed for conversion or extraction operations must be free of side-effects and shall in partiular not throw exceptions, because the invocation of the function is part of the pattern matching process.

\begin{lstlisting}
auto f= [](const string& str)-> option<int> {
  char* p = nullptr;
  auto result = strtol(str.c_str(), &p, 10);
  if (p && *p == '\0') return result;
  return {};
};
receive (
  on(f) >> [](int i) {
    // case 1, conversion successful
  },
  on_arg_match >> [](const string& str) {
    // case 2, str is not an integer
  }
);
\end{lstlisting}

The lambda function \lstinline^f^ is a \lstinline^string^ $\Rightarrow$ \lstinline^int^ projection, but rather than returning an integer, it returns \lstinline^option<int>^.
An empty \lstinline^option^ indicates that a value does not have a valid mapping to an integer.
Functors used as projection or extration must take exactly one argument and must return a value.
The types for the pattern are deduced from the functor's signature.
If the functor returns an \lstinline^option<T>^, then \lstinline^T^ is deduced.

Our DSL-based approach to pattern matching has more syntactic noise than a native support within the languages itself, i.e., compared to functional programming languages such as Haskell or Erlang.
However, our approach uses only ISO C++11 facilities, does not rely on brittle macro definitions, and after all adds little -if any- runtime overhead by using expression templates \cite{v-et-95}.

%% file: part5.tex
\section{Message Passing}
\label{Sec:MessagePassing}

By supporting thread-like and event-based actors, \libcppa has to support blocking as well as asynchronous operations.
By using a behavior-based approach, we are able to provide an asynchronous API without inversion of control.

\subsection{Receiving Messages}

We provide a blocking API to receive messages for threaded and context-switching actors.
The blocking function \lstinline^receive^ sequentially iterates over all elements in the mailbox beginning with the first.
It takes a partial function that is applied to the elements in the mailbox until an element was matched.
An actor calling \lstinline^receive^ is blocked until it successfully dequeued a message from its mailbox or an optional timeout occurs.

\begin{lstlisting}
receive (
  on... >> // ...
  after(std::chrono::seconds(1)) >> // timeout
);
\end{lstlisting}

The code snippet above illustrates the use of \lstinline^receive^.
Note that the partial function passed to \lstinline^receive^ is a temporary object at runtime.
Hence, using receive inside a loop would cause creation of a new partial function on each iteration.
To avoid re-creation of temporal objects per loop iteration, \libcppa provides three predefined receive loops, \lstinline^receive_loop^, \lstinline^receive_for^, and \lstinline^do_receive(...).until^, to provide a more efficient but yet convenient way of defining receive loops.

Due to the callback-based nature of event-based message handling, we have to provide a non-blocking API, too.
Hence, we provide a behavior-based API, which enables event-like message handling without inversion of control \cite{ho-jmlc-06}.

An event-based actor uses \lstinline^become^ to set its behavior.
The given behavior is then executed until it is replaced by another call to \lstinline^become^ or the actor finishes execution.
Class-based actor definitions simply subtype \lstinline^event_based_actor^ and must implement the pure virtual member function \lstinline^init^.
An implementation of \lstinline^init^ shall set an initial behavior by using \lstinline^become^.

\begin{lstlisting}
struct printer : event_based_actor {
  void init() { become (
    others() >> [] {
      cout << to_string(self->last_received())
           << endl;
    }
  ); }
};
\end{lstlisting}

Another way to implement event-based actors is provided by the class \lstinline^sb_actor^ (``State-Based Actor'').
This base class calls \lstinline^become(init_state)^ in its \lstinline^init^ member function.
Hence, a subclass must only provide a member of type \lstinline^behavior^ named \lstinline^init_state^.

\begin{lstlisting}
struct printer : sb_actor<printer> {
  behavior init_state = (
    others() >> [] {
      cout << to_string(self->last_received())
           << endl;
    }
  );
};
\end{lstlisting}

Note that \lstinline^sb_actor^ uses the Curiously Recurring Template Pattern \cite{c-crtp-95}, why the derived class must be given as template parameter.
This technique allows \lstinline^sb_actor^ to access the \lstinline^init_state^ member of a derived class.

The following example illustrates a more advanced state-based actor that implements a stack with a fixed maximum number of elements.
Since this example uses non-static member initialization, it might not compile with some currently available compilers.

\begin{lstlisting}
struct fixed_stack : sb_actor<fixed_stack> {
  static constexpr size_t max_size = 10;
  std::vector<int> data;
  behavior empty = (
    on(atom("push"), arg_match) >>
    [=](int what) {
      data.push_back(what);
      become(filled);
    },
    on(atom("pop")) >> [=]() {
      reply(atom("failure"));
    }
  );
  behavior filled = (
    on(atom("push"), arg_match) >>
    [=](int what) {
      data.push_back(what);
      if (data.size() == max_size)
        become(full);
    },
    on(atom("pop")) >> [=]() {
      reply(atom("ok"), data.back());
      data.pop_back();
      if (data.empty()) become(empty);
    }
  );
  behavior full = (
    on<atom("push"),int>() >> []() {
      // discard value
    },
    on(atom("pop")) >> [=]() {
      reply(atom("ok"), data.back());
      data.pop_back();
      become(filled);
    }
  );
  behavior& init_state = empty;
};
\end{lstlisting}

Nesting receives in an event-based actor is slightly more difficult compared to context-switching or thread-mapped actors, since \lstinline^become^ does not block.
An actor has to set a new behavior calling \lstinline^become^ with the \lstinline^keep_behavior^ policy to wait for the required message and then return to the previous behavior by using \lstinline^unbecome^, as shown in the example below.
An event-based actor finishes execution with normal exit reason if the behavior stack is empty after calling \lstinline^unbecome^.
The default policy of \lstinline^become^ is \lstinline^discard_behavior^ that causes an actor to override its current behavior.
The policy flag must be the first argument of \lstinline^become^.

\begin{lstlisting}
// receives {int, float} sequences
struct testee : sb_actor<testee> {
  behavior init_state = (
    on<int>() >> [=](int value1) {
      become (
        // the keep_behavior policy stores
        // the current behavior on the
        // behavior stack to be able to
        // return to this behavior later on
        // by calling unbecome()
        keep_behavior,
        on<float>() >> [=](float value2) {
          cout << value1 << " => "
               << value2 << endl;
          unbecome();
        }
      );
    }
  );
};
\end{lstlisting}

\subsection{Sending Messages}

The default way of passing a message to another actor is provided by the function \lstinline^send^, which models asynchronous communication.
However, we also provide a \lstinline^sync_send^ function that returns a handle to the response message.
Synchronous response messages can be received using this handle only and are not visible otherwise by receive operations.
This allows actors to identify response messages unambiguously.
Furthermore, whenever receiving of a synchronous response message times out, the message is dropped even if it is received later on.

The functions \lstinline^receive_response^ and \lstinline^handle_response^ can be used to receive response messages, as shown in the following example.

\begin{lstlisting}
// replies to atom("get") with a string
auto testee = spawn<testee_impl>();
// blocking API
auto future = sync_send(testee, atom("get"));
receive_response (future) (
  on_arg_match >> [&](const string& str) {
    // handle str
  },
  after(chrono::seconds(30)) >> [&]() {
    // handle error
  }
);
// event-based actor API (nonblocking)
auto future = sync_send(testee, atom("get"));
handle_response (future) (
  on_arg_match >> [=](const string& str) {
    // handle str
  },
  after(chrono::seconds(30)) >> [=]() {
    // handle error
  }
);
\end{lstlisting}

The function \lstinline^receive_response^ is similar to \lstinline^receive^, i.e., it blocks the calling actor until either a response message arrived or a timeout occures.

Similar to \lstinline^become^, the function \lstinline^handle_response^ is part of the event-based API and is used as ``one-shot handler''.
The behavior passed to \lstinline^handle_response^ is executed \textit{once} and the actor automatically returns to its previous behavior afterwards.
It is possible to `stack' multiple \lstinline^handle_response^ calls.
Each response handler is executed once and then automatically discarded.

In both cases, the behavior definition of the response handler requires a timeout, because a non-replying actor would always cause a deadlock otherwise.

Often times, an actor sends a synchronous message and then wants to wait for the response immediately afterwards.
In this case, using either \lstinline^handle_response^ or \lstinline ^receive_response^ is quite verbose.
Therefore, the returned handle provides the two member functions \lstinline^then^ and \lstinline^await^.
Using \lstinline^then^ is equal to using \lstinline^handle_response^, wheres \lstinline^await^ corresponds to \lstinline^receive_response^, as illustrated by the following example.

\begin{lstlisting}
// blocking API
sync_send(testee, atom("get")).await(
  on_arg_match >> [&](const string& str) {
    // handle str
  },
  after(chrono::seconds(30)) >> [&]() {
    // handle error
  }
);
// event-based actor API (nonblocking)
sync_send(testee, atom("get")).then(
  on_arg_match >> [=](const string& str) {
    // handle str
  },
  after(chrono::seconds(30)) >> [=]() {
    // handle error
  }
);
\end{lstlisting}

%% file: part6.tex
\section{Mailbox Implementation}
\label{Sec:Mailbox}

\begin{figure}[htp]
 \centering
 \includegraphics[width=0.85\columnwidth]{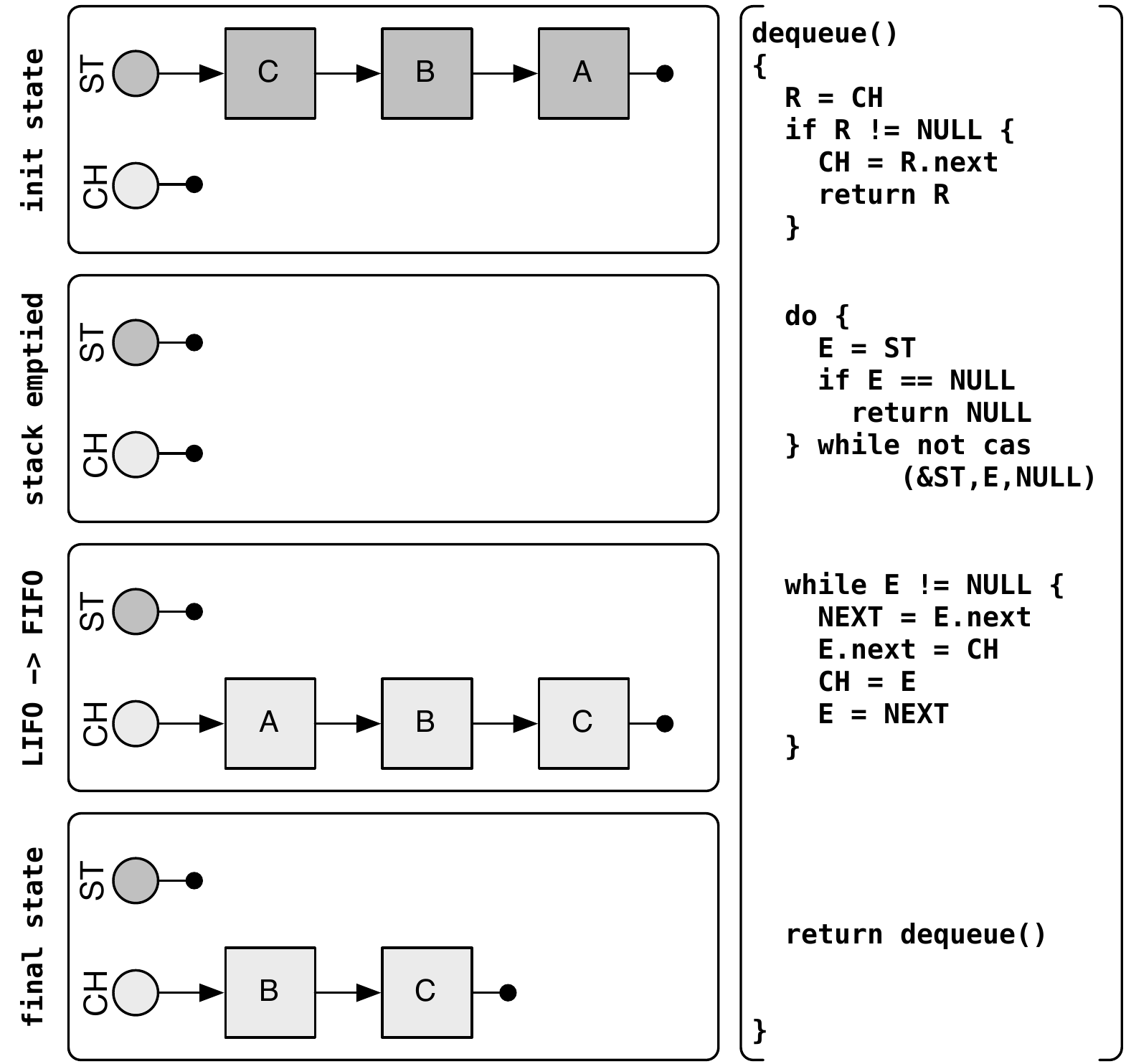}
 \label{Fig::CachedStackPop}
 \caption{Dequeue operation in a cached stack (ST = \textbf{S}tack \textbf{T}ail, CH = \textbf{C}ache \textbf{H}ead)}
\end{figure}

The message queue or \textit{mailbox} implementation is a critical component of any message passing systems.
All messages sent to an actor are delivered to its mailbox, which acts as a shared resource whenever an actor receives messages from multiple senders in parallel.
Thus, the overall system~performance, foremost its scalability depends
significantly on the selected algorithm.

A mailbox is a single-reader-many-writer queue.
Everyone is allowed to enqueue a message to a mailbox, but only the owning actor is allowed to dequeue a message.
Hence, the dequeue operation does not need to support parallel access.
We have combined a lock-free stack implementation with a FIFO ordered queue as internal cache.
A lock-free stack can be implemented using a single atomic compare-and-swap (CAS) operation.
It does not suffer from the so called $ABA$ problem of concurrent access that can corrupt states in
CAS-based systems \cite{ibmc-iseap-83} as the enqueue operation only needs to manipulate the \textit{tail} pointer.
However, without reordering the dequeue operation would have to traverse the (LIFO-sorted)
stack in order to find the oldest element.

Figure \ref{Fig::CachedStackPop} shows the dequeue operation of our mailbox implementation.
It always dequeues elements from the FIFO ordered cache (CH).
The stack (ST) is emptied and its elements are moved in reverse order to the cache, after the cache was drained.
Emptying the stack can be done by a single CAS operation as it only needs to set ST to NULL.

Our mailbox has complexity $O(1)$ for enqueue operations while the dequeue operation has an average of $O(1)$ but a worst case of $O(n)$, where $n$\ is the current number of messages in the stack. 
Concurrent access to the cached stack is reduced to a minimum and both enqueueing and dequeueing perform only a single CAS operation.

\subsection{Performance for N:1 Communication}
\label{Sec:Mailbox:Performance}

To measure the efficiency of our mailbox implementation, we compare the runtime behavior of our software with common implementations of the actor model.
Erlang and Scala are currently the most relevant languages for actor programming.
We use their implementation as reference for concurrent computation.
For Scala, we consider its standard library as well as the well-established third party library Akka~\cite{ti-a-12}.
In detail, our benchmarks are based on the following implementations of the actor model.

\begin{description}

\item[cppa]
    \libcppa applying event-based message processing.

\item[erlang]
    \erlang in version 5.9.2.

\item[scala actors]
    \scala with the event-based actor implementation of the standard library.

\item[scala akka]
    \scala with the \akka library in version 2.0.3

\end{description}

\begin{figure}
\centering%
\includegraphics[width=0.9\columnwidth]{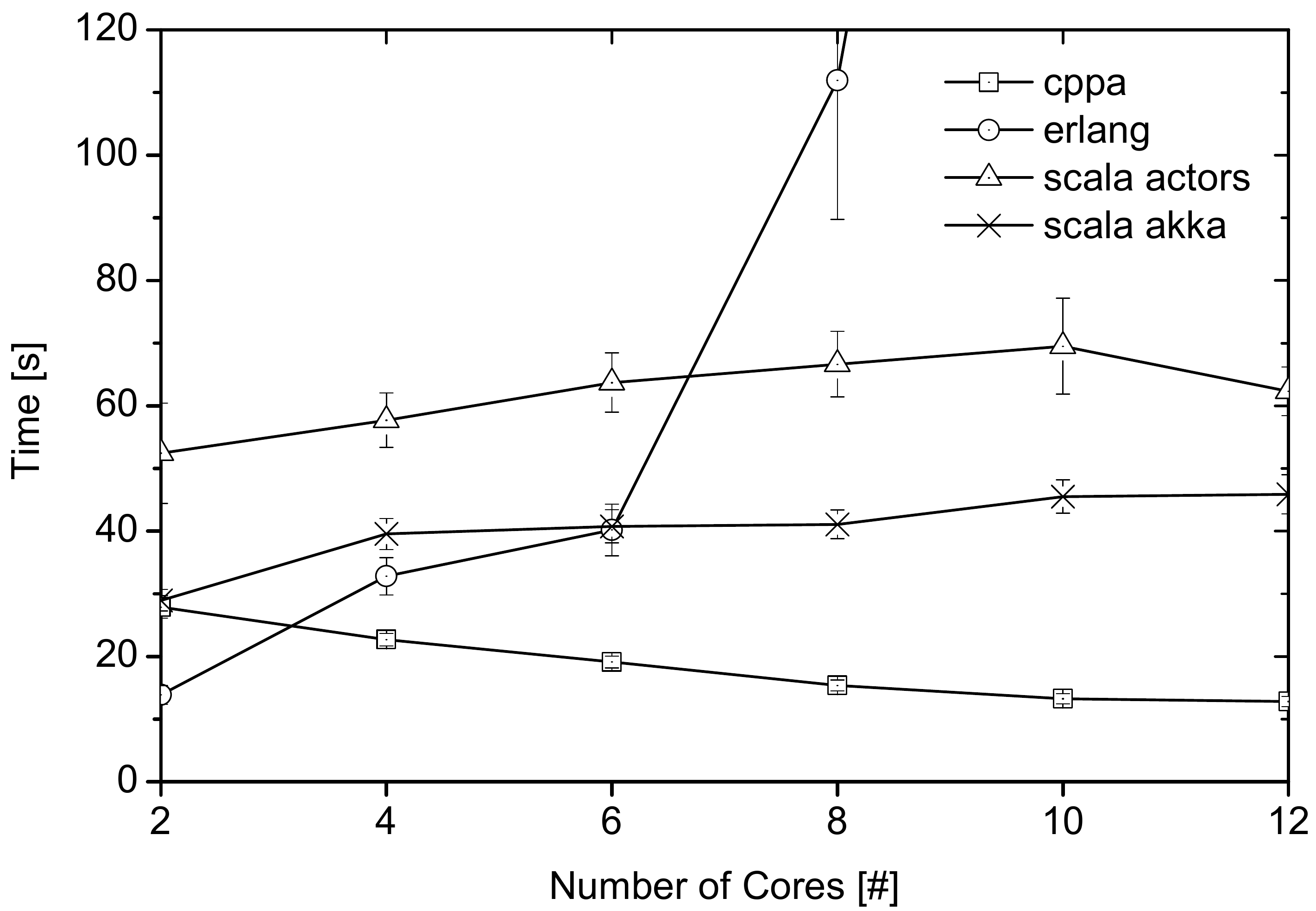}%
\label{Fig::Mailbox::Time}%
\caption{Mailbox performance in N:1 communication scenario}
\end{figure}

\begin{figure*}%
\centering%
\subfigure[Sending and processing time]{\includegraphics[width=0.9\columnwidth]{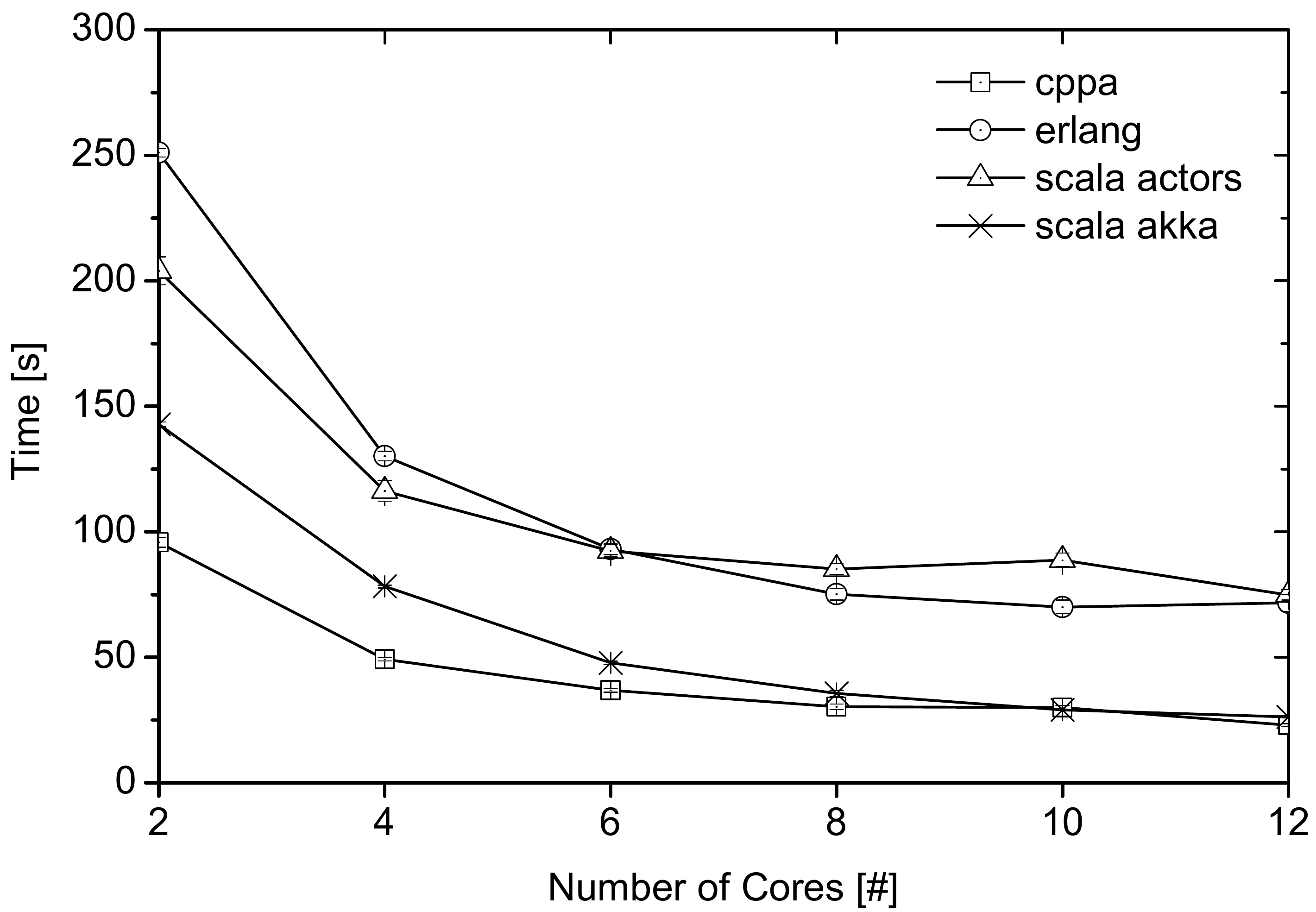}\label{Fig::MixedCase::Time}}%
\hspace{1.5cm}
\subfigure[Memory consumption]{\includegraphics[width=0.9\columnwidth]{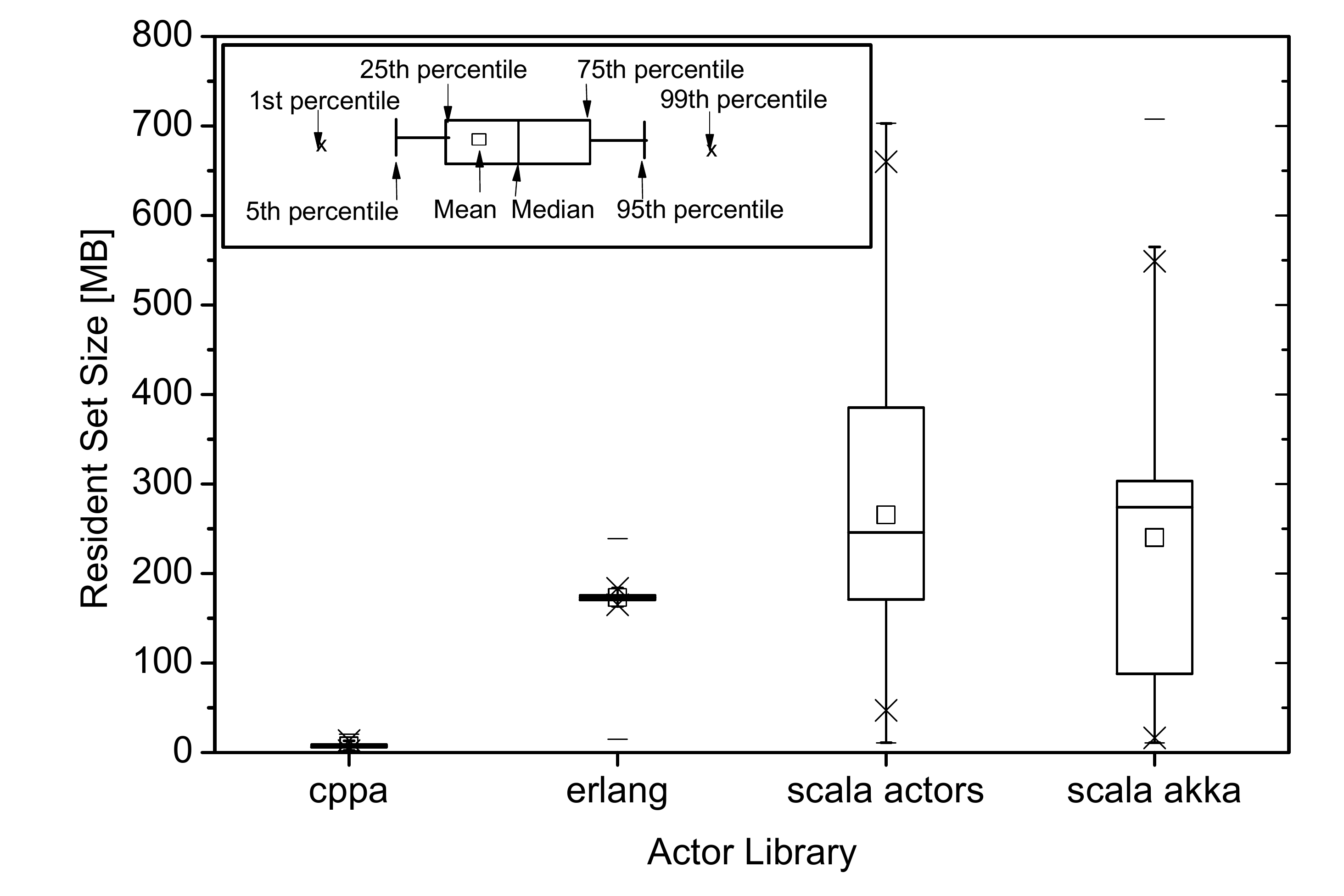}\label{Fig::MixedCase::Memory}}%
\caption{Performance in a mixed scenario with additional work load}\label{Fig::MixedCasePerformance}%
\end{figure*}

\noindent \libcppa has been compiled with optimization level O4 of  GNU \cpp compiler version 4.7.2.
\scala version 2.9.1 runs on a JVM configured with 4~GB of RAM.

We used 20 threads sending 1,000,000 messages each, except for \erlang, which has no threading library.
In \erlang, we spawned 20 actors instead.
The minimal runtime of this benchmark is the time the receiving actor needs to process the 20,000,000 messages and the overhead of passing the messages to the mailbox.
More hardware concurrency leads to higher synchronization between the sending threads, since the mailbox acts as a shared resource.
Furthermore, the workers in the thread pool are synchronized by a job queue in the implementations using a cooperative scheduling, which can be a concurrency bottleneck as well if actors perform short tasks and often need re-scheduling.

Figure~\ref{Fig::Mailbox::Time} visualizes the time needed for the application to send and process the 20,000,000 messages as a function of available CPU cores.
The ideal behavior is a decreasing curve, which reaches a global minimum given by the time the receiving actor needs to consume all messages one by one.
For visibility reasons, we cut the y-axis at 120~s.

The processing overhead increases significantly for \erlang in case of more than 6~cores and directly correlates with the degree of  concurrency.
On average, it requires up to 700~seconds on 12~cores.
\libcppa attains the best behavior -- a constantly decreasing curve -- as its performance gain on additional hardware concurrency clearly outweighs the increasing synchronization overhead.
Akka slightly outperforms \scala~Actors, but has a steadily increasing curve, whereas \scala~Actors reaches it maximum on ten cores and accelerates again on twelve cores.
The results indicate that our mailbox implementation using the cached stack algorithm as well as our synchronization protocol among worker threads scale very well.

%% file: part7.tex
\section{Mixed Use Case Performance}
\label{Sec:PerformanceEvaluation}

In this section, we consider a realistic use case including a mixture of operations under severe work load to evaluate the performance of \libcppa.
We again used the setup described in Section \ref{Sec:Mailbox:Performance} for this benchmark, but also measure the memory consumption to examine both runtime and resource efficiency.
The program creates a simple multi-ring topology with a fixed number of actors per ring.
A token with an initial value of 10,000 is passed along the ring and decremented each round.
A client that receives the token forwards it to its neighbor and terminates whenever the value of the token is 0.
Thus, we continuously create and terminate actors, which process a total of 50,000,000 messages.

We also create one worker per ring that calculates the prime factors of 28,350,160,440,309,881, i.e., 329,545,133 and 86,028,157, to add numerical work load.
It is worth noting that this operation is independent of any other actor and does not involve messages.
The calculation requires about two seconds in our loop-based \cpp implementation.
Our tail-recursive \scala implementation of the prime factorization operates at the same speed, whereas \erlang needs almost seven seconds to finish the calculation.

In total, we create 20 rings. Each ring consists of
49 \lstinline^chain_link^ actors and one \lstinline^master^. The
\lstinline^master^ re-creates the terminated actors five times. Each
\lstinline^master^ thus spawns a total of 245 actors. Additionally, there
is one message collector and one worker per master. The message collector
waits until it receives the result of 100 (20$\cdot$5) prime factorizations
and a \textit{done} message from each master. Overall 4,921 actors are created, but no more than 1021 actors run concurrently. 

The following pseudo code illustrates the implemented algorithm.

\begin{lstlisting}
master(Worker, Collector):
 5 times:
  Next = self
  49 times: Next = spawn(chain_link, Next)
  Next ! {token, 10000}
  Worker ! {calc, 28350160440309881}
  Done = false
  while not Done:
   receive:
    {token, X} =>
     if (X > 0): Next ! {token, X-1}
     else: Done = true
 Collector ! {master_done}

chain_link(Next):
 receive:
  {token, N} =>
   Next ! {token, N}
   if (N > 0) chain_link(Next)

worker(Collector):
 receive:
  {calc, X} =>
   Collector ! {result, fact(X)}
\end{lstlisting}

Figure~\ref{Fig::MixedCase::Time} shows the runtime behaviour 
as a function of available CPU cores.
An ideal characteristic would halve the runtime when doubling the number of cores.
\erlang exhibits an almost linear speed-up, while \libcppa and Akka decrease monotonically but non-linearly.
\scala~Actors increase in runtime after a minimum at eight cores, but accelerates again on twelve cores.
\erlang performs well considering our previous observation that its prime factorization is more than three times slower. The efficient scheduling of \erlang, which is the only implementation in our measurement that performs \emph{preemptive} scheduling, perfectly utilizes hardware concurrency up to ten core, when it reaches its global minimum.
\akka is significantly faster than the standard library implementations of \scala but is slightly outperformed by \libcppa.

Figure~\ref{Fig::MixedCase::Memory} shows the memory consumption during the mixed scenario.
Both Erlang and \libcppa have a very constant and thus predictable memory footprint.
As it seems, Erlang's virtual machine pre-allocates about 180 MB and never needs to allocate additional memory during the runtime.
\libcppa uses only a fraction of the memory compared to all other implementations and accounts for the benchmark's characteristics -- constant number of actors and messages on average -- as memory is released as soon as possible.
Akka as well as \scala~Actors exhibit a very unpredictable memory consumption, which cannot be explained by the benchmark characteristics, with an average memory consumption between 250 and 300~MB, and peaks above 600~MB.

%% file: part8.tex
\section{Limitations Induced by C++}
\label{Sec:Limitations}

Pattern matching has proven useful for a lot of different scenarios and is not limited to message handling in actor systems.
We think our approach pushes the possibilities to emulate pattern matching using template metaprogramming to the limit.
Yet, a library-based approach can only bring some of the power of pattern matching to C++ and is far less elegant than built-in facilities for instance known from Haskell or Scala.
Furthermore, splitting the pattern part from the definition of the variables that holds matching parts makes our approach more verbose -- and thus harder to read -- than it could be with proper language support.

Due to the lack of pure functions in C++, we cannot check at compile time whether projection and conversion functions fulfill our requirement of not having side-effects and not throwing exceptions.
Relying on conventions rather than enforcing requirements is a serious shortcoming and can lead to subtle bugs that are very difficult to find, even for expert programmers.

Furthermore, lambdas passed to \lstinline^become^ must not have references to the enclosing scope.
Since \lstinline^become^ always returns immediately, all references are guaranteed to cause undefined behavior.
Again, we have no chance of checking and enforcing this requirement, since the type of the lambda expression does not exhibit any information other than its signature.
This problem is probably difficult to address in the language specification itself, but standardized attributes to annotate this kind of requirement would allow static analyzers to find related bugs with relative ease.

The adaption of language facilities previously found in functional programming languages like type inference, lambda expressions, and partial function application provided by \lstinline^std::bind^, raised the expressive power of C++.
In fact, \libcppa would most likely not be possible in the C++03 standard.
In our opinion, the community should stay on this path of including more and more ``functional'' features, such as pattern matching and maybe monads, as they make asynchronous APIs very simple and potentially straightforward to implement.
Although new features increase the overall size of the language, it allows developers to write cleaner and shorter code.

%% file: part9.tex
\section{Future Work}
\label{Sec:FutureWork}

One of the main strengths of the actor model is its abstraction over heterogenous environments.
With the increasing importance of GPGPU programming -- GPUs are known to outperform
CPUs in computationally intensive applications --, it is a natural next step to provide facilities to create GPU compliant actors in \libcppa.
Since a GPU has its own memory, code and
data are transferred to the GPU before executing an algorithm and the results
must be read back after the computation is done.
This management could be done by \libcppa, since this workflow fits very well to the message passing paradigm.
An GPU actor could define its behavior based on patterns, but would have to provide an additional GPGPU implementation, e.g., by using OpenCL.
Such actors would be scheduled on the GPU rather than on the cooperatively used thread pool if an appropriate graphics card was found.
Executing actors on a GPU would enable \libcppa to address high-performance computation applications based on the actor model as well.

As for synchronous messaging, we want to relax the requirement of being forced to use timeouts.
Instead, the runtime system could send an error message if the receiving actor has already exited.

Overall, we think \libcppa can serve as a good tool for C++11 developers.
As an additional future direction, we will focus on supporting ARM platforms and embedded systems by optimizing \libcppa to be even more resource efficient.
In this way, we hope to contribute our share to  ease development of native, distributed  applications in both high-end and low-end computing.